\documentclass[12pt,a4paper]{revtex4}
\usepackage{amsmath}
\usepackage{amssymb}
\usepackage{bm}
\usepackage{graphicx}
\usepackage{wrapfig}
\usepackage{wrapft}
\begin{document}

\title{%
Short-time critical dynamics of the three-dimensional systems with long-range correlated disorder%
}

\author{
Vladimir V. \textsc{Prudnikov}$^{1,}$\footnote{E-mail: prudnikv@univer.omsk.su},
Pavel V. \textsc{Prudnikov}$^{1}$,
Bo \textsc{Zheng}$^{2}$,\\
Sergei V. \textsc{Dorofeev}$^{1}$ and
Vyacheslav Yu. \textsc{Kolesnikov}$^{1}$,%
}

\affiliation{%
$^1$Dept. of Theoretical Physics, Omsk State University, Omsk 644077, Russia\\
$^2$Physics Dept., Zhejiang University, Hangzhou 310027, PR of China
}

\begin{abstract}
Monte Carlo simulations of the short-time dynamic behavior are reported
for three-dimensional Ising and XY models with long-range correlated
disorder at criticality, in the case corresponding to linear defects.
The static and dynamic critical exponents are determined for systems
starting separately from ordered and disordered initial states.
The obtained values of the exponents are in a
good agreement with results of the field-theoretic description of the
critical behavior of these models in the two-loop approximation and
with our results of Monte Carlo simulations of three-dimensional Ising
model in equilibrium state.
\end{abstract}

\maketitle

\section{Introduction\label{sec:1}}

The investigation of critical behavior of disordered systems remains
one of the main problems in condensed matter physics and excites a great
interest, because all real solids contain structural defects.
The structural disorder breaks the translational symmetry of the crystal
and thus greatly complicates the theoretical description of the material.
The influence of disorder is particularly important near critical point
where behavior of a system is characterized by anomalous large response
on any even weak perturbation. In most investigations consideration has
been restricted to the case of point-like uncorrelated defects \cite{1}.
However, the non-idealities of structure cannot be modeled by simple
uncorrelated defects only. Solids often contain defects of a more complex
structure: linear dislocations, planar grain boundaries, clusters of
point-like defects, and so on.

Different models of structural disorder have arisen as an attempt to
describe such complicated defects. In this paper we concentrate on
model of Weinrib and Halperin (WH) \cite{2} with the so-called
long-range correlated disorder when pair correlation function for
point-like defects $g({\bf x}-{\bf y})$ falls off with distance
as a power law $g({\bf x}-{\bf y})\sim |{\bf x - y}|^{-a}$.
Weinrib and Halperin showed that for $a\geq d$ long-range correlations are
irrelevant and the usual short-range Harris criterion \cite{3} $2 - d\nu_0 = \alpha_0 > 0$
of the effect of point-like uncorrelated defects is realized, where $d$ is
the spatial dimension, and $\nu_0$ and $\alpha_0$ are the correlation-length
and the specific-heat exponents of the pure system. For $a < d$ the extended
criterion $2 - a\nu_0 > 0$ of the effect of disorder on the critical
behavior was established. As a result, a wider class of disordered systems,
not only the three-dimensional diluted Ising model with point-like uncorrelated
defects, can be characterized by a new type of critical behavior. So, for
$a < d$  a new long-range (LR) disorder stable fixed point (FP) of the
renormalization group recursion relations for systems with a number of
components of the order parameter $m \geq 2$ was discovered. The critical
exponents were calculated in the one-loop approximation using a double
expansion in $\varepsilon = 4 - d \ll 1$ and $\delta = 4 - a \ll 1$.
The correlation-length exponent was evaluated in this linear approximation
as $\nu=2/a$ and it was argued that this scaling relation is exact and also
holds in higher order approximation. In the case $m = 1$ the accidental
degeneracy of the recursion relations in the one-loop approximation did
not permit to find LR disorder stable FP. Korzhenevskii {\it et al}
\cite{4} proved the existence of the LR disorder stable FP for the
one-component WH model and also found characteristics of this type of
critical behavior.

Ballesteros and Parisi \cite{5} have studied by Monte Carlo means the
critical behavior in equilibrium of the 3D site-diluted Ising model
with LR spatially correlated disorder, in the $a=2$ case corresponding
to linear defects. They have computed the critical exponents of these
systems with the use of the finite-size scaling techniques and found that a
$\nu$ value is compatible with the analytical predictions $\nu=2/a$.

However, numerous investigations of pure and disordered systems performed
with the use of the field-theoretic approach show that the predictions made
in the one-loop approximation, especially based on the $\varepsilon$ -
expansion, can differ strongly from the real critical behavior \cite{6,7,8,9}.
Therefore, the results for WH model with LR correlated defects received
based on the $\varepsilon, \delta$ - expansion \cite{2,4,10,11,12} was
questioned in our paper \cite{13}, where a renormalization analysis of
scaling functions was carried out directly for the 3D systems in the
two-loop approximation with the values of $a$ in the range
$2\leq a \leq 3$, and the FPs corresponding to stability of various
types of critical behavior were identified. The static and dynamic critical
exponents in the two-loop approximation were calculated with the use of
the Pade-Borel summation technique. The results obtained in \cite{13}
essentially differ from the results evaluated by a double $\varepsilon, \delta$ -
expansion. The comparison of calculated the exponent $\nu$ values and ratio $2/a$
showed the violation of the relation $\nu = 2/a$, supposed in \cite{2} as exact.

The models with LR-correlated quenched defects have both theoretical
interest due to the possibility of predicting new types of critical
behavior in disordered systems and experimental interest due to the
possibility of realizing LR-correlated defects in the orientational
glasses \cite{14}, polymers \cite{15}, and disordered solids containing
fractal-like defects \cite{4} or dislocations near the sample
surface \cite{16}.

To shed light on the reason of discrepancy between the results of Monte Carlo
simulation of the 3D Ising model with LR-correlated disorder
\cite{5}, in the $a=2$ case, and the results of our renormalization group
description of this model \cite{13}, we have computed by the short-time
dynamics method \cite{17,18} the static and dynamic critical exponents
for site-diluted 3D Ising and XY models with the linear defects of
random orientation in a sample.

In the following section, we introduce a site-diluted 3D Ising model
with the linear defects and scaling relations for the short-time critical
dynamics. In Section~\ref{sec:3}, we give results of critical temperature determination
for 3D Ising model with the linear defects for case with spin concentration
$p=0.8$. We analyze the critical short-time dynamics in Ising systems
starting separately from ordered and disordered initial states. Critical
exponents obtained under these two conditions with the use of the corrections
to scaling are compared. Also, in Section~\ref{sec:3} the results
of measurements of the critical characteristics for 3D Ising model in
equilibrium state are presented in comparison with results of the short-time
dynamics method. In Section~\ref{sec:4} the results of Monte Carlo studies of
critical behavior of 3D XY-model with linear defects for the same spin
concentration $p=0.8$ are considered. Our main conclusions are discussed
in Section~\ref{sec:5}.

\section{Model and Methods\label{sec:2}}

We have considered the following Ising model Hamiltonian defined in a cubic
lattice of linear size $L$ with periodic boundary conditions:
\begin{eqnarray}
H =-J\sum_{\langle i,j\rangle}p_i p_j S_i S_j,
\end{eqnarray}
where the sum is extended to the nearest neighbors, $S_i=\pm 1$ are the
usual spin variables, and the $p_i$ are quenched random variables
($p_i=1$, when the site $i$ is occupied by spin, and $p_i=0$, when the site
is empty), with LR spatial correlation. An actual $p_i$ set will
be called a sample from now on. We have studied the next way
to introduce the correlation between the $p_i$ variables for WH model with
$a=2$, corresponding to linear defects. We start with a filled cubic lattice
and remove lines of spins until we get the fixed spin concentration $p$ in
the sample. We remove lines along the coordinate axes only to preserve
the lattice symmetries and equalize the probability of removal for all
the lattice points. This model was referred in \cite{5} as the model with
non-Gaussian distribution noise and characterized by the isotropic
impurity-impurity pair correlation function decays for large $r$ as
$g(r)\sim 1/r^2$. In contrast to \cite{5} we put a condition
of linear defects disjointness on their distribution in a sample, whereas
in \cite{5} the possibility of linear defects intersection is not discarded.
The physical grounds for this condition are connected with fact that in real
materials dislocations as linear defects are distributed uniformly in
macroscopic sample with probability of their intersection close to zero.
The condition of linear defects disjointness corresponds to WH model since
the intersection of linear defects being taken into consideration results in
additional vertices of interaction which are absent in the effective
Hamiltonian of WH model.

In this paper we have investigated systems with the spin concentration
$p=0.8$. We have considered the cubic lattices with linear sizes
$L$ from 16 to 128. The Metropolis algorithm has been used in simulations.
We consider only the dynamic evolution of systems described by the model A
in the classification of Hohenberg and Halperin \cite{19}.
The Metropolis Monte Carlo scheme of simulation with the dynamics of a
single-spin flips reflects the dynamics of model A and enables us
to compare the obtained dynamical critical exponent $z$ with the results
of our renormalization group description of critical dynamics of
this model \cite{13} having LR-disorder.

A lot of results have been recently obtained concerning the critical
dynamical behavior of statistical models \cite{17,18} in the macroscopic
short-time regime.
This kind of investigation was motivated by analytical and numerical
results contained in the papers of Janssen {\it et al} \cite{20} and
Huse \cite{21}. Important is that extra critical exponents should be
introduced to describe the dependence of the scaling behavior for
thermodynamic and correlation functions on the initial conditions.
According to the argument of Janssen, Schaub and Schmittman \cite{20}
obtained with the renormalization group method, one may expect a
generalized scaling relation for the $k$-th moment of the magnetization
\begin{eqnarray} \label{eq:kmagn}
M^{(k)}(t,\tau,L,m_0) &=& b^{-k\beta/\nu} M^{(k)}\left(b^{-z}t,b^{1/\nu}\tau,b^{-1}L,b^{x_0}m_0\right),
\end{eqnarray}
is realized after a time scale $t_{mic}$ which is large enough in microscopic
sense but still very small in macroscopic sense. In (\ref{eq:kmagn}) $\beta$, $\nu$
are the well-known static critical exponents and $z$ is the dynamic exponent,
while the new independent exponent $x_0$ is the scaling dimension of the
initial magnetization $m_0$, $\tau=(T-T_c)/T_c$ is the reduced temperature.

Since the system is in the early stage of the evolution the correlation
length is still small and finite size problems are nearly absent.
Therefore we generally consider $L$ large enough and skip this argument.
We now choose the scaling factor $b=t^{1/z}$ so that the main $t$-dependence
on the right is cancelled. Expanding the scaling form (\ref{eq:kmagn}) for $k=1$ with
respect to the small quantity $t^{\,x_0/z}m_0$, one obtains
\begin{eqnarray} \label{eq:theta}
M(t,\tau,m_0) \sim m_0 t^{\theta} F(t^{1/\nu z}\tau,t^{x_0/z}m_0)
              = m_0 t^{\theta}(1+at^{1/\nu z}\tau)+O(\tau^2, m_0^2),
\end{eqnarray}
where $\theta=(x_0 - \beta/\nu)/z$ has been introduced. For $\tau=0$ and
small enough $t$ and $m_0$ the scaling dependence for magnetization (\ref{eq:theta}) takes
the form $M(t) \sim m_0 t^{\theta}$. For almost all statistical systems studied
up to now \cite{17,18,22}, the exponent $\theta$ is positive, i.e., the magnetization
undergoes surprisingly a critical initial increase. The time scale of this
increase is $t_0 \sim m_0^{-z/x_0}$. However, in the limit of $m_0$ the time
scale goes to infinity. Hence the initial condition can leave its trace even
in the long-time regime.

If $\tau \neq 0$, the power law behavior is modified by the scaling function
$F(t^{1/\nu z} \tau)$ with corrections to the simple power law, which will be depended
on the sign of $\tau$. Therefore, simulation of the system for temperatures
near the critical point allows to obtain the time dependent magnetization with
non-perfect power behavior, and the critical temperature $T_c$ can be determined
by interpolation.

Other two interesting observables in short-time dynamics are the second moment
of magnetization $M^{(2)}(t)$ and the auto-correlation function
\begin{equation}
A(t)= \frac{1}{L^d}\left<\sum_{i}S_i(t)S_i(0)\right>.
\end{equation}
As the spatial correlation length in the beginning of the time evolution is small,
for a finite system of dimension $d$ with lattice size $L$ the second moment
$M^{(2)}(t,L) \sim L^{-d}$. Combining this with the result of the scaling form (\ref{eq:kmagn})
for $\tau=0$ and $b=t^{1/z}$, one obtains
\begin{equation} \label{eq:c2}
M^{(2)}(t) \sim t^{\, -2\beta/\nu z} M^{(2)}\left(1,t^{-1/z}L\right) \sim t^{\displaystyle \, c_2}, \qquad
       c_2 = \left(d-2\frac{\beta}{\nu}\right)\frac{1}{z}.
\end{equation}
Furthermore, careful scaling analysis shows that the auto-correlation also decays by a
power law \cite{23}
\begin{equation} \label{eq:ca}
A(t) \sim t^{\displaystyle \, -c_a}, \qquad
 c_a = \frac{d}{z}-\theta.
\end{equation}
Thus, the investigation of the short-time evolution of system from a high-temperature
initial state with $m_0=0$ allows to determine the dynamic exponent $z$, the ratio
of static exponents $\beta/\nu$ and a new independent critical exponent $\theta$.

Till now a completely disordered initial state has been considered as starting
point, i.e., a state of very high temperature. The question arises how a completely
ordered initial state evolves, when heated up suddenly to the critical temperature.
In the scaling form (\ref{eq:kmagn}) one can skip besides $L$, also the argument $m_0=1$
\begin{equation}
M^{(k)}(t,\tau)=b^{-k\beta/\nu}M^{(k)}\left(b^{-z}t,b^{1/\nu}\tau \right).
\end{equation}
The system is simulated numerically by starting with a completely ordered state,
whose evaluation is measured at or near the critical temperature.
The quantities measured are $M(t)$, $M^{(2)}(t)$. With $b=t^{1/z}$ one avoids the main
$t$-dependence in $M^{(k)}(t)$, and for $k=1$ one has
\begin{eqnarray} \label{eq:moder}
M(t,\tau)=t^{-\beta/\nu z}M(1,t^{1/\nu z}\tau)
         = t^{-\beta/\nu z}\left(1+at^{1/\nu z}\tau+O(\tau^2)\right).
\end{eqnarray}
For $\tau=0$ the magnetization decays by a power law $M(t)\sim t^{-\beta/\nu z}$.
If $\tau \neq 0$, the power law behavior is modified by the scaling function
$M(1,t^{1/\nu z} \tau)$. From this fact, the critical temperature $T_c$ and
the critical exponent $\beta/\nu z$ can be determined.

We must note, that the short-time dynamic method in part of critical evolution
description of system starting from the ordered initial state is essentially
the same as the non-equilibrium relaxation method proposed by N.~Ito in
\cite{24} for critical behavior study of three-dimensional pure Ising model.
At present, this method was extended by N.~Ito to non-equilibrium relaxation
study of Ising spin glass models \cite{25}, Kosterlitz-Thouless phase
transition \cite{26} and fully frustrated XY models in two dimension \cite{27}.

From scaling form (\ref{eq:moder}) the power law of time dependence for the logarifmic
derivative of the magnetization can be obtained in the next form
\begin{equation} \label{eq:logderm}
\left.\partial_{\tau} ln  M(t,\tau) \right|_{\tau=0} \sim t^{1/\nu z},
\end{equation}
which allows to determine the ratio $1/\nu z$. On basis of the magnetization and its
second moment the time dependent Binder cumulant
\begin{equation} \label{eq:bcum}
U_2(t)= \frac{M^{(2)}}{(M)^2} - 1 \sim t^{d/z}
\end{equation}
is defined. From its slope one can directly measure the dynamic exponent $z$.
Consequently, from an investigation of the system relaxation from ordered initial state
with $m_0=1$ the dynamic exponent $z$ and the static exponents $\beta$ and $\nu$
can be determined and their values can be compared with results of simulation of system
behavior from disordered initial state with $m_0=0$.

\section{Measurements of the critical temperature and critical exponents
for 3D Ising model}\label{sec:3}

We have performed simulations on three-dimensional cubic lattices with linear sizes $L$
from 16 to 128, starting either from an ordered state or from a high-temperature state
with zero or small initial magnetization.
We would like to mention that measurements starting from from a completely ordered
state with the spins oriented in the same direction ($m_0=1$) are more favorable,
since they are much less affected by fluctuations, because the quantities measured
are rather big in contrast to those from a random start with $m_0=1$.
Therefore, for careful determination of the critical temperature and critical exponents
for 3D Ising model with linear defects we begin to investigate the relaxation of
this model from a completely ordered initial state.

\subsection{Evolution from an ordered state}\label{sec:3a}

Initial configurations for systems with the spin concentration $p=0.8$ and
randomly distributed quenched linear defects were generated numerically.
Starting from those initial configurations, the system was updated with Metropolis
algorithm. Simulation have been performed up to $t=1000$.
We measured the time evolution of the magnetization
\begin{equation} \label{eq:m}
M(t)=\frac{1}{N_{spin}}\left[\left<\sum_{i}p_iS_i(t) \right>\right]
\end{equation}
and the second moment
\begin{equation} \label{eq:m2}
M^{(2)}(t)=\frac{1}{N^2_{spin}}\left[\left<\left(\sum_{i}p_iS_i(t)\right)^2 \right>\right],
\end{equation}
which also allow to calculate the time dependent Binder cumulant $U_2(t)$ (\ref{eq:bcum}).
The angle brackets in (\ref{eq:m}) and (\ref{eq:m2}) denote the statistical averaging and the
square brackets are for averaging over the different impurity configurations.

\begin{figure}
\centerline{
\includegraphics[width=0.75\textwidth]{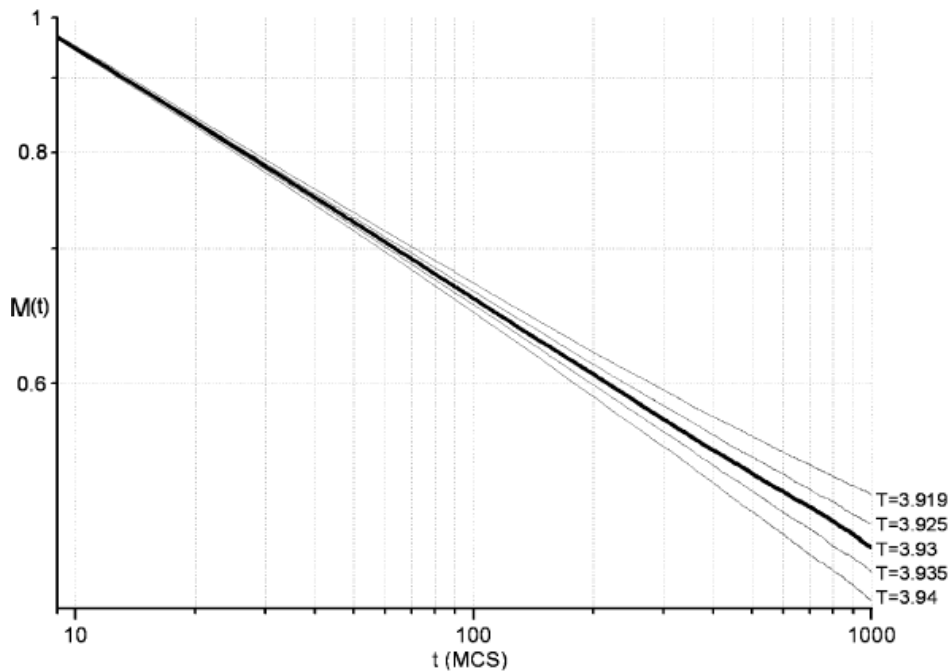}}
\caption{ \label{fig:1} Time evolution of the magnetization $M(t)$ for $L=128$
and for different values of the temperature $T$. }
\end{figure}

In Fig.~\ref{fig:1} the magnetization $M(t)$ for samples with linear size $L=128$ at
$T=3.919,$ 3.925, 3.930, 3.935 and 3.940 is plotted in log-log scale.
The resulting curves in Fig.~\ref{fig:1} have been obtained by averaging over 3000
samples with different linear defects configurations. We have determined
the critical temperature $T_c = 3.930(2)$ from best fitting of these
curves by power law.

The critical temperature determined in \cite{5} for
the same system with spin concentration $p=0.8$ in the non-Gaussian case is
$T_c = 3.8891(2)$. This difference of the critical temperature values
shows that different principles of distribution of linear defects are
the reason of discrepancy between the results obtained in \cite{5}
by Monte Carlo simulation of the 3D Ising model with LR-correlated
disorder, and results in renormalization group description of this model
\cite{13}.

In order to check-up the critical temperature value independently,
we have carried out in equilibrium the calculation of Binder cumulant $U_4$,
defined as
\begin{equation}
U_4 = \frac{1}{2}\left( 3 - \frac{[\left<M^4\right>]\phantom{^2}}{[\left<M^2\right>]^2}\right),
\end{equation}
and the correlation length \cite{28}
\begin{eqnarray}
\xi&=&\frac{1}{2\sin{(\pi/L)}}\sqrt{\frac{\chi}{F}-1\,}, \\
\chi&=&\frac{1}{N_{spin}}[\langle M^2\rangle], \\
F&=&\frac{1}{N_{spin}}[\langle\Phi\rangle],  \\
\Phi&=&\frac{1}{3}\sum_{n=1}^{3}\left(\left|\sum_j{p_jS_j \exp\left(\frac{2\pi ix_{n,j}}{L}\right)}\right|^2\right),
\end{eqnarray}
where $(x_{1,j},x_{2,j},x_{3,j})$ are coordinates of $j$-th site of lattice.

\begin{figure}
\includegraphics[width=0.45\textwidth]{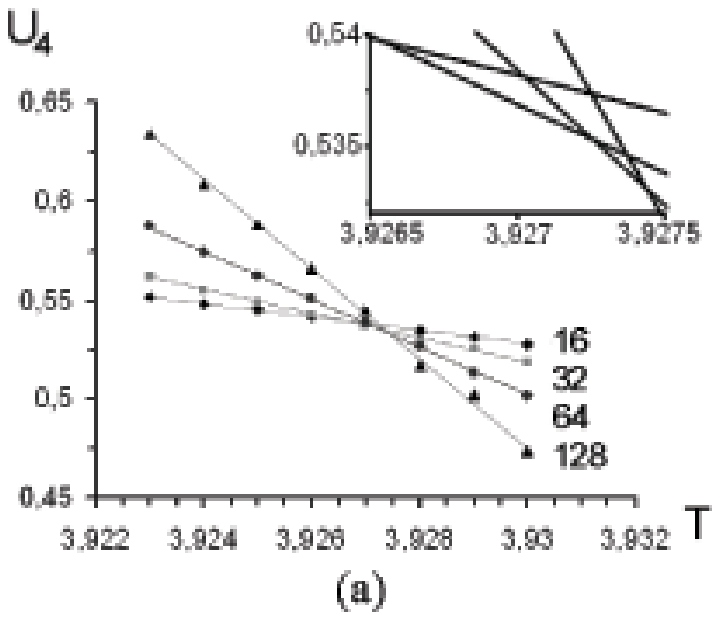}
\includegraphics[width=0.45\textwidth]{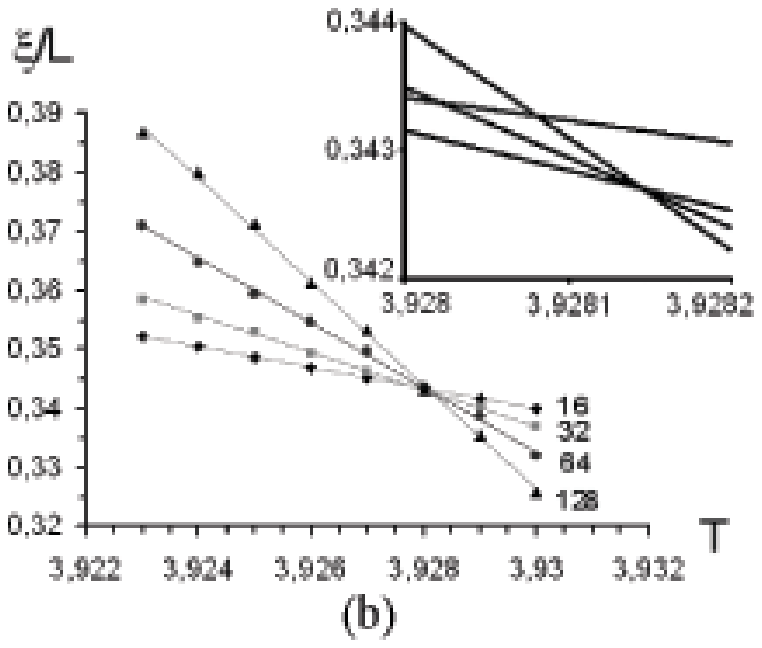}
\caption{ \label{fig:2} Binder cumulant $U_4(T,L)$ (a) and ratio $\xi/L$ (b) as a function
of T for lattices with different sizes $L$. }
\includegraphics[width=0.45\textwidth]{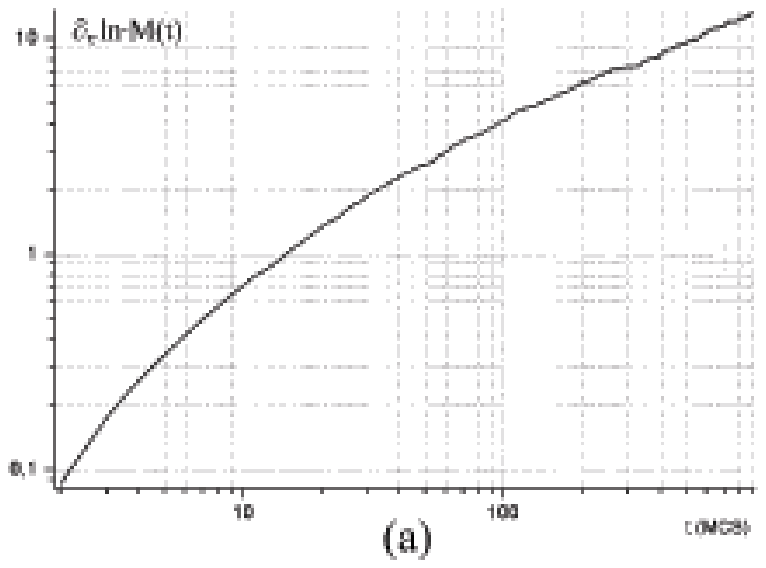}
\includegraphics[width=0.45\textwidth]{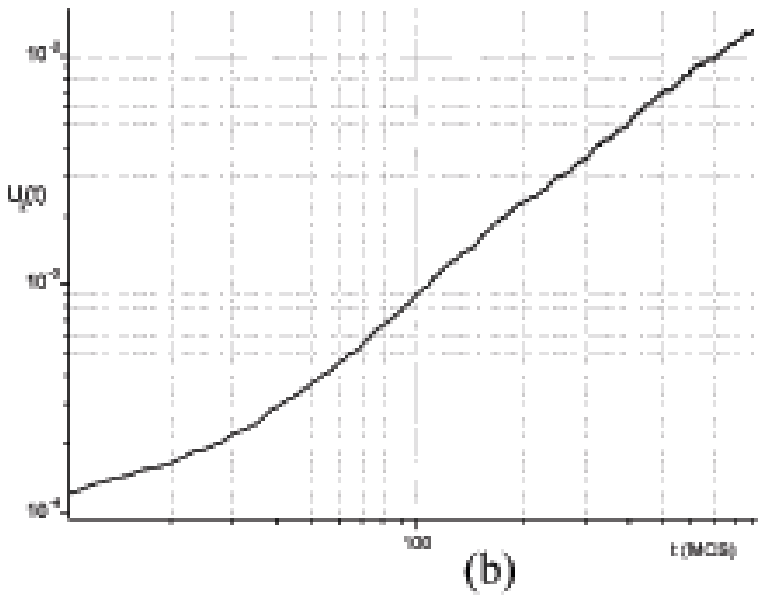}
\caption{ \label{fig:3} Time evolution of logarithmic derivative of the magnetization
$\left.\partial_{\tau} ln  M(t,\tau) \right|_{\tau=0}$ (a) and the cumulant $U_2(t)$ (b)
for $L=128$ at the critical temperature $T_c=3.9281$.}
\end{figure}

The cumulant $U_4 (L,T)$ has a scaling form
\begin{equation}
U_4(L,T) = u\left(L^{1/\nu}(T-T_c)\right).
\end{equation}
The scaling dependence of the cumulant makes it possible to determine the
critical temperature $T_c$ from the coordinate of the points
of intersections of the curves specifying the temperature dependence
$U_4 (L,T)$ for different $L$. In Fig.~\ref{fig:2}a the computed curves of $U_4(L,T)$
are presented for lattices with sizes $L$ from 16 to 128. As a result it was
determined that the critical temperature is $T_c=3.9275(5)$.
In this case for simulations we have used the Wolff single-cluster algorithm
with elementary MCS step as 5 cluster flips. We discard 10000 MCS for
equilibration and then measure after every MCS with averaging over 100000
MCS. The results have been averaged over 15000 different samples for
lattices with sizes $L=16, 32$ and over 10000 samples for lattices
with sizes $L=64, 128$.

The crossing of $\xi/L$ was introduced as a convenient method for
calculating of $T_c$ in \cite{29}.
In Fig.~\ref{fig:2}b the computed curves of temperature dependence of ratio
$\xi/L$ are presented for lattices with the same sizes, the coordinate
of the points of intersections of which also gives the critical temperature
$T_c=3.9281(1)$. This value of $T_c$ we selected as the best for subsequent
investigations of the Ising model.

Also, we have determined the temperature of intersection
of the curves specifying the temperature dependence cumulants
$U_4 (L,T)$ for $L=16$ and $L=32$ with the use of linear defects distribution
in samples as in \cite{5} with the possibility of their intersection.
Computation gives $T_c(L)=3.8884(6)$ in this case which corresponds to
the results in \cite{5} but differs from $T_c(L)=3.9185(5)$ obtained with
the use of condition of linear defects disjointness for lattices
with the same sizes.
Turning back to short-time dynamics method, we note that the exponent
$1/\nu z$ can be determined from relation (\ref{eq:logderm}) if we differentiate
$ln M(t,\tau)$ with respect to $\tau$.
\begin{wrapfigure}[31]{r}{7cm}   
\vspace*{-2mm}\includegraphics[width=0.45\textwidth]{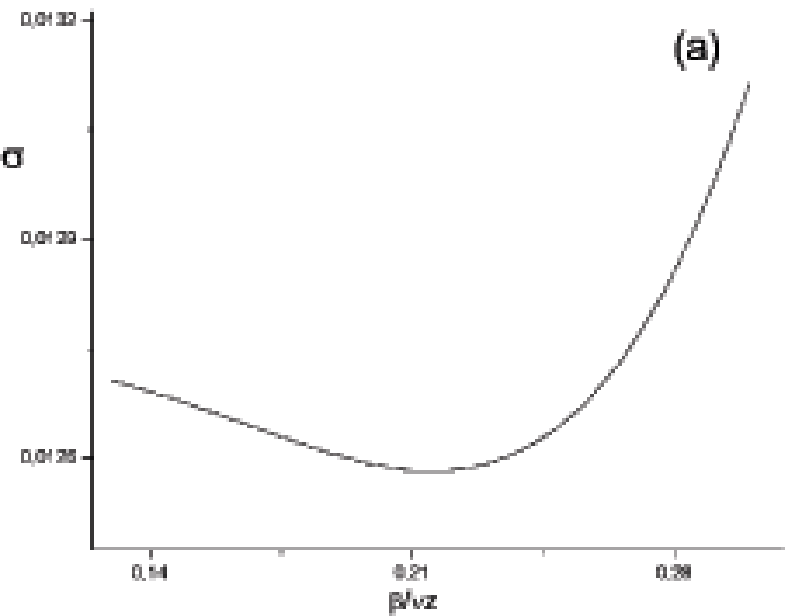}\\[2mm]
\includegraphics[width=0.45\textwidth]{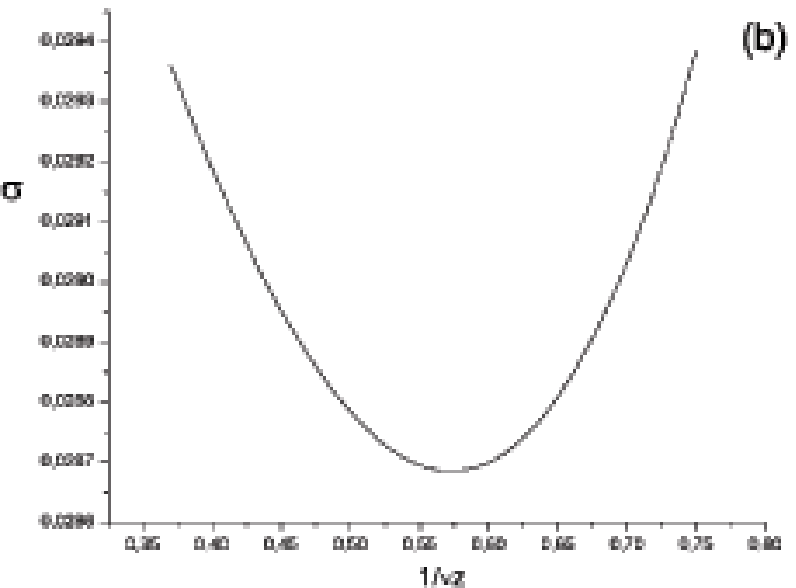}\\[2mm]
\includegraphics[width=0.45\textwidth]{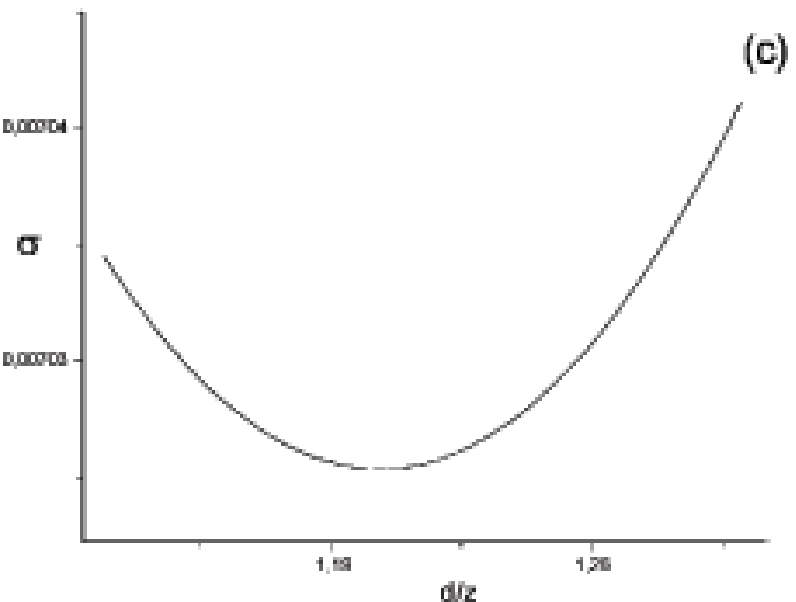}\\[2mm]
\vspace*{-15mm}\\
\caption{ \label{fig:4} Dependence of the mean square errors $\sigma$ of the fits for the
magnetization (a), logarifmic derivative of the magnetization (b), and
the cumulant (c) as a function of the exponents $\beta/\nu z$,
$1/\nu z$, and $d/z$ for $\omega=0.8$. }
\end{wrapfigure}
The dynamic exponent $z$ can be determined from analysis of time dependent
Binder cumulant $U_2(t)$ (\ref{eq:bcum}) for $\tau=0$.
In Fig.~\ref{fig:3} the logarithmic derivative of the magnetization
$\left.\partial_{\tau} ln  M(t,\tau) \right|_{\tau=0}$ with respect to $\tau$ (Fig.~\ref{fig:3}a) and
the cumulant $U_2(t)$ (Fig.~\ref{fig:3}b) for samples with linear size $L=128$ at $T_c=3.9281$
are plotted in log-log scale. The $\left.\partial_{\tau} ln  M(t,\tau) \right|_{\tau=0}$
have been obtained from a quadratic interpolation between the three curves
of time evolution of the magnetization for the temperatures
$T= 3.9250$, $3.9281$, $3.9310$ and taken at the critical temperature $T_c=3.9281$.
The resulting curves have been obtained by averaging over 3000 samples.

We have analyzed the time dependence of the cumulant $U_2(t)$ and clarified
that in the time interval [50,150] the $U_2(t)$ is best fitted by power law
with the dynamic exponent $z \simeq 2.02$, corresponding to the pure Ising
model \cite{30,31}, and the linear defects are developed for $t>400$ MCS only.
An analysis of the $U_2(t)$ slope measured in the interval [500,900] shows
that the exponent $d/z=1.173(12)$ which gives $z=2.558(26)$.
We have taken into account these dynamic crossover effects
for analysis of the time dependence of magnetization and its derivative.
So, the slope of magnetization and its derivative over the interval [450,900]
provides the exponets $\beta/\nu z =0.230(2)$ and $1/\nu z=0.517(12)$ which
give $\nu=0.746(19)$ and $\beta=0.445(10)$.

In the next stage, we have considered the corrections to the scaling in order
to obtain accurate values of the critical exponents. We have applied the following
expression for the observable~$X(t)$:
\begin{eqnarray}
X(t) \sim t^{\Delta}\left(1+A_x t^{-\omega/z}\right),
\end{eqnarray}
where $\omega$ is a well-known exponent of corrections to scaling.
This expression reflects the scaling transformation in the critical range
of time-dependent corrections to scaling in the form of $t^{-\omega/z}$
to the usual form of corrections to scaling $\tau^{\omega \nu}$ in equilibrium
state for time t comparable with the order parameter relaxation time
$t_r \sim \xi^z \Omega (k\xi)$ \cite{17}.
Field-theoretic estimate of the $\omega$ value gives $\omega \simeq 0.80$
in the two-loop approximation \cite{15}. Monte Carlo study of Ballesteros and Parisi
\cite{5} shows that $\omega \simeq 1.0$.

\begin{table}[t]
\caption{\label{tab:1} Values of the exponents $\beta/\nu z$, $1/\nu z$, $d/z$, and
minimal values of the mean square errors $\sigma$ in fits for different
values of the exponent $\omega$}
\begin{center}
\begin{tabular}{lllllll} \hline \hline
 \multicolumn{1}{c}{$\omega$} & \multicolumn{1}{c}{$\beta/\nu z$}
& \multicolumn{1}{c}{$\sigma$} & \multicolumn{1}{c}{$1/\nu z$}
& \multicolumn{1}{c}{$\sigma$} & \multicolumn{1}{c}{$d/z$} & \multicolumn{1}{c}{$\sigma$} \\ \hline
 0.7 & 0.2112 & 0.0100 & 0.556 & 0.0053 & 1.183 & 0.0100 \\
 0.8 & 0.2096 & 0.0088 & 0.559 & 0.0049 & 1.205 & 0.0100 \\
 0.9 & 0.2101 & 0.0093 & 0.553 & 0.0070 & 1.213 & 0.0099 \\
 1.0 & 0.2090 & 0.0095 & 0.558 & 0.0072 & 1.227 & 0.0098 \\ \hline
\end{tabular}
\end{center}
\end{table}

To analyze our simulation date we have used the linear approximation
of the $(X t^{-\Delta})$ on $t^{-\omega/z}$ with the changing values of the
exponent $\Delta$ and the exponent $\omega$ from the interval [0.7,1.0].
Then, we have investigated the dependence of the mean square errors $\sigma$
of this fitting procedure for the function $X t^{-\Delta}(t^{-\omega/z})$ on
the changing $\Delta$ and $\omega$. In Fig.~\ref{fig:4} we plot the $\sigma$ for
the magnetization (Fig.~\ref{fig:4}a), logarifmic derivative of the magnetization
(Fig.~\ref{fig:4}b), and the cumulant (Fig.~\ref{fig:4}c) as a function of the exponents
$\beta/\nu z$, $1/\nu z$, and $d/z$ for $\omega=0.8$. Minimum of $\sigma$
determines the exponents $z$, $\nu$, and $\beta$ for every $\omega$.
In Table~\ref{tab:1} we present the computed values of the exponents $\beta/\nu z$,
$1/\nu z$, and $d/z$, and minimal values of the mean square errors $\sigma$
in these fits for values of the exponent $\omega=0.7, 0.8, 0.9, 1.0$.
We see that the values of $\beta/\nu z$, $1/\nu z$, and $d/z$ are weakly
dependent on the change of the exponent $\omega$ in the interval [0.7,1.0], but
the $\omega=0.8$ is preferable because it gives the best fit for the
magnetization and the logarifmic derivative of the magnetization dates.
Finally, for the $\omega=0.8$ we find the following values of the exponents
\begin{equation}
\begin{array}{rl}
& z = 2.489 \pm 0.021,  \\
& \nu = 0.719 \pm 0.022,   \\
& \beta = 0.375 \pm 0.045.
\end{array}
\end{equation}

It is interesting to compare these values of exponents with those obtained in
\cite{13} with the use of the field-theoretic approach
\begin{equation}
\begin{array}{rl}
& z = 2.495,  \\
& \nu = 0.716,   \\
& \beta = 0.350,
\end{array}
\end{equation}
which demonstrate a very good agreement with each other, but show an
essential difference from Monte Carlo results of Ballesteros and Parisi
\cite{5} with $\nu = 1.009(13)$ and $\beta=0.526(15)$.

\subsection{Evolution from a disordered state}\label{sec:3b}

\begin{figure}
\centerline{
\includegraphics[width=0.45\textwidth]{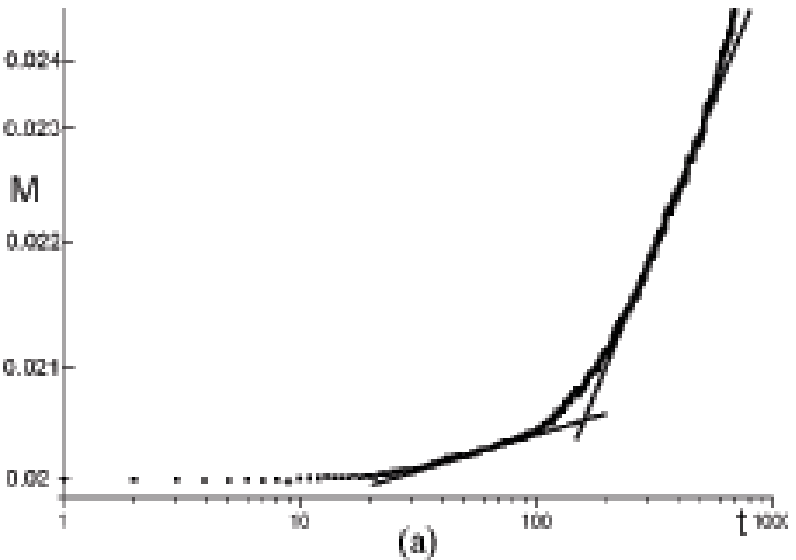}
\includegraphics[width=0.45\textwidth]{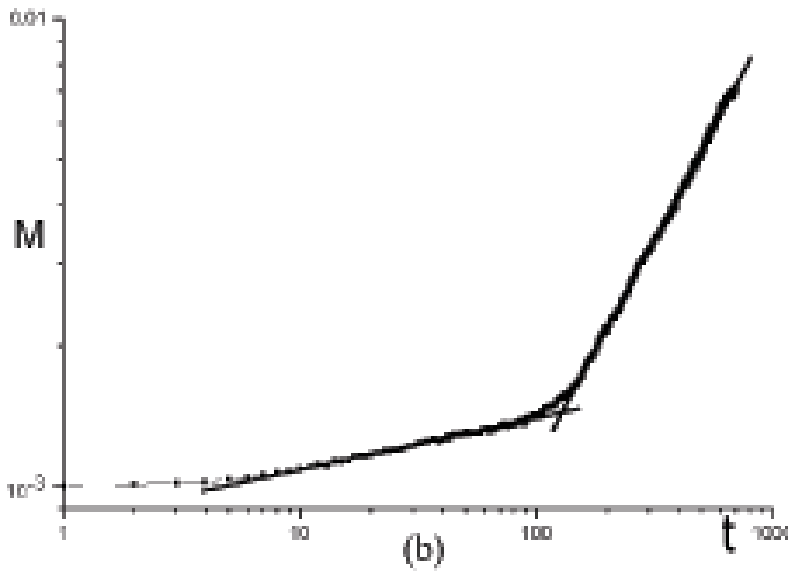}}
\caption{ \label{fig:5} Time evolution of the magnetization $M(t)$
for $L=128$ with the initial magnetization $m_0=0.02$ (a)
and $m_0=0.001$ (b) at the critical temperature $T_c=3.9281$. }
\centerline{
\includegraphics[width=0.45\textwidth]{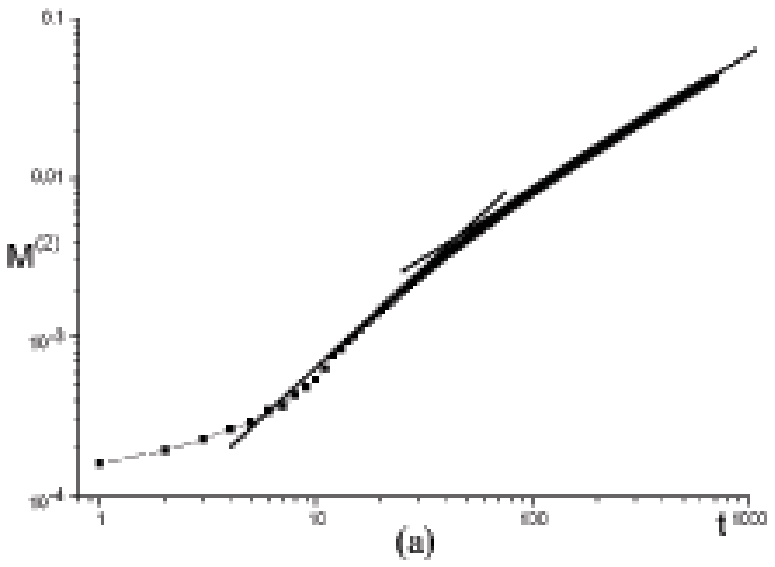}
\includegraphics[width=0.45\textwidth]{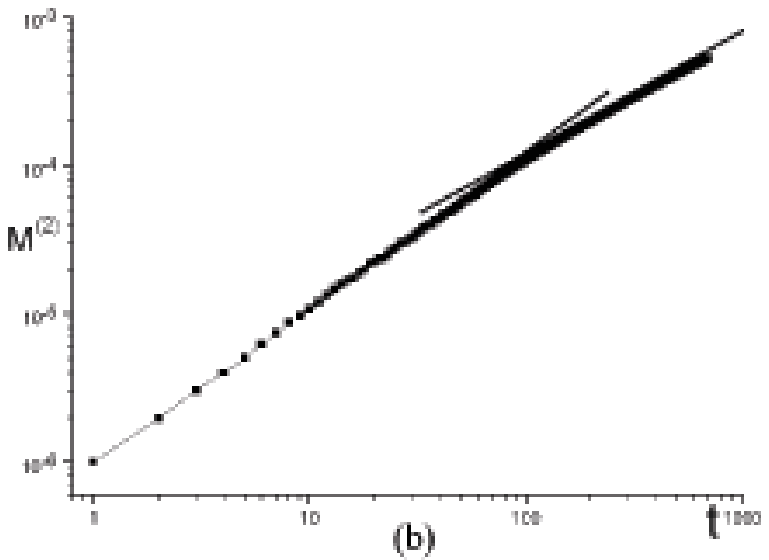}}
\caption{ \label{fig:6} Time evolution of the second moment $M^{(2)}$
for $L=128$ with the initial magnetization $m_0=0.02$ (a)
and $m_0=0.001$ (b) at the critical temperature $T_c=3.9281$. }
\end{figure}

We have also performed simulations of evolution of the system with linear defects
on the largest lattice with $L=128$, starting from a disordered state with
small initial magnetizations $m_0=0.02$ and $m_0=0.001$ at the critical point.
The initial magnetization has been prepared by flipping in an ordered state
a definite number of spins at randomly chosen sites  in order to get the desired
small value of $m_0$. In accordance with Section~\ref{sec:2}, a generalized dynamic
scaling predicts in this case a power law evolution for the magnetization $M(t)$,
the second moment $M^{(2)}(t)$ and the auto-correlation $A(t)$ in the short-dynamic region.

\begin{wrapfigure}[11]{r}{6.6cm}   
\includegraphics[width=0.45\textwidth]{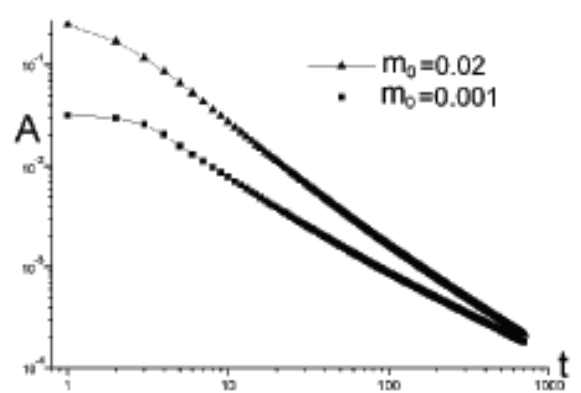}
\caption{ \label{fig:7} Time evolution of the autocorrelation $A(t)$
for $L=128$ with different initial magnetization values $m_0$
at the critical temperature $T_c=3.9281$. }
\end{wrapfigure}

In Fig.~\ref{fig:5}, \ref{fig:6} and \ref{fig:7} we show the obtained curves for $M(t)$ (Fig.~\ref{fig:5}),
$M^{(2)}(t)$ (Fig.~\ref{fig:6}) and $A(t)$ (Fig.~\ref{fig:7}), which are plotted
in log-log scale up to $t=700$. These curves were resulted by averaging
over 3000 different samples with 25 runs for each sample.
From Fig.~\ref{fig:5} we can see an initial increase of the magnetization,
which is a very prominent phenomenon in the short-time critical
dynamics \cite{17,18}. But in contrast to dynamics of the pure systems
\cite{17}, we can observe the crossover from dynamics of the pure system
on early times of the magnetization evolution up to $t=70$
to dynamics of the disordered system with the influence of
linear defects in the time interval [100,650].
The same crossover
phenomena were observed in evolution of the second moment
$M^{(2)}(t)$ and the autocorrelation $A(t)$.
In result of linear approximation of these curves in the both
time intervals we obtained the values of the exponents $\theta$, $c_2$
and $c_a$ in accordance with relations in (\ref{eq:theta}), (\ref{eq:c2}) and (\ref{eq:ca}) for
initial states with $m_0=0.02$ and $m_0=0.001$ (Table~\ref{tab:2}).
The final values of these exponents and also the critical exponents
$z$, $\beta/\nu$ and $x_0$ were obtained by extrapolation to $m_0=0$.
In Table~\ref{tab:2} we compare the values of these exponents
with values of corresponding exponents for the pure Ising model \cite{18}
and theoretical field description (TFD) results for system with linear defects \cite{13}.
The obtained values quite well agree with results of simulation from an ordered
state with $m_0=1$ and with results from \cite{13} and show that
LR-correlated defects lead to faster increasing of the magnetization
in the short-time dynamic regime in compare with the pure system.

\begin{table}[t]
\caption{\label{tab:2} Values of the critical exponents obtained in present work for
evolution from disordered initial states with different $m_0$ and
extrapolated to $m_0=0$ and corresponding exponents
for pure systems \cite{17} and from \cite{13}}
\begin{tabular}{lllllll} \hline \hline
            & \multicolumn{1}{c}{$\theta$}    & \multicolumn{1}{c}{$c_2$}
                        & \multicolumn{1}{c}{$c_a$}       & \multicolumn{1}{c}{$z$}
                        & \multicolumn{1}{c}{$\beta/\nu$} & \multicolumn{1}{c}{$x_0$}    \\ \hline
 $t \in [10,70]$         &&&&&&                                                                               \\ \hline
 $m_0=0,02$             & 0.086(12) & 0.964(28)  & 1.384(26)   &            &           &            \\
 $m_0=0,001$            & 0.099(9)  & 0.973(19)  & 1.364(23)   &            &           &            \\
 $m_0=0$            & 0.101(10) & 0.975(23)  & 1.363(26)   & 2.049(27)  & 0.501(27) & 0.708(34)  \\       \hline
 $t \in [100,650]$        &&&&&&                                                                              \\   \hline
 $m_0=0,02$             & 0.152(12)  & 0.812(21)  & 1.103(16)  &            &           &            \\
 $m_0=0,001$            & 0.149(10)  & 0.804(19)  & 1.047(12)  &            &           &            \\
 $m_0=0$            & 0.149(11)  & 0.801(20)  & 1.043(14)  & 2.517(32)  & 0.492(28) & 0.867(37)  \\
 TFD  \cite{13}         &            &            &            & 2.495      & 0.489     &            \\
 pure \,\cite{17}   & 0.108(2)   & 0.970(11)  & 1.362(19)  & 2.041(18)  & 0.510(14) & 0.730(25)  \\ \hline
\end{tabular}
\end{table}

\subsection{Measurements of the critical characteristics
in equilibrium state}\label{sec:3c}

With the aim to verify the short-time dynamics me\-thod and the results obtained
we also carried out the study of the critical behavior of 3D Ising model with the
linear defects of random orientation by traditional Monte Carlo simulation
methods in equilibrium state. For simulations we have used the Wolf
single-cluster algorithm. We have computed for the critical temperature
$T_c = 3.9281(1)$ the values of different thermodynamic and correlation
functions in equilibrium state such as the magnetization, susceptibility,
correlation length, heat capacity, and Binder cumulant $U_4$ for lattices
with sizes $L$ from 16 to 128 for the same spin concentration $p=0.8$.
The use of well-known scaling critical dependences for these thermodynamic
and correlation functions with taking into consideration the finite size
scaling corrections
\textfloatsep=5mm
\begin{figure}[!t]
\begin{center}
\includegraphics[width=0.45\textwidth]{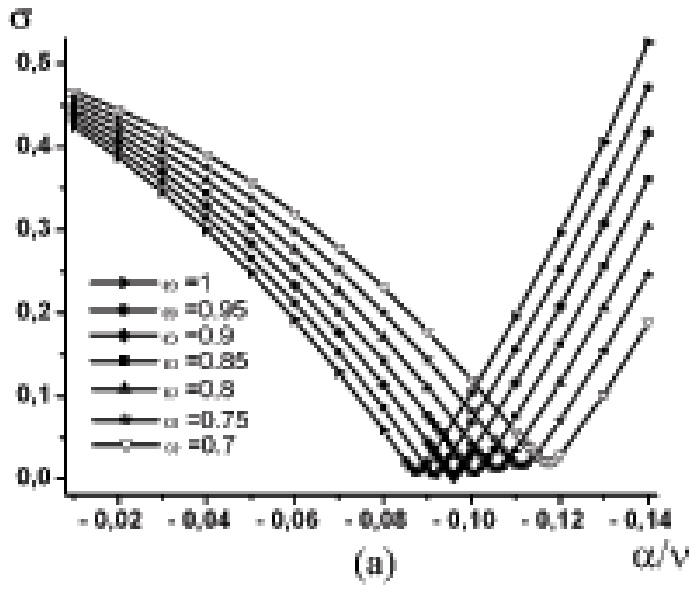}
\includegraphics[width=0.45\textwidth]{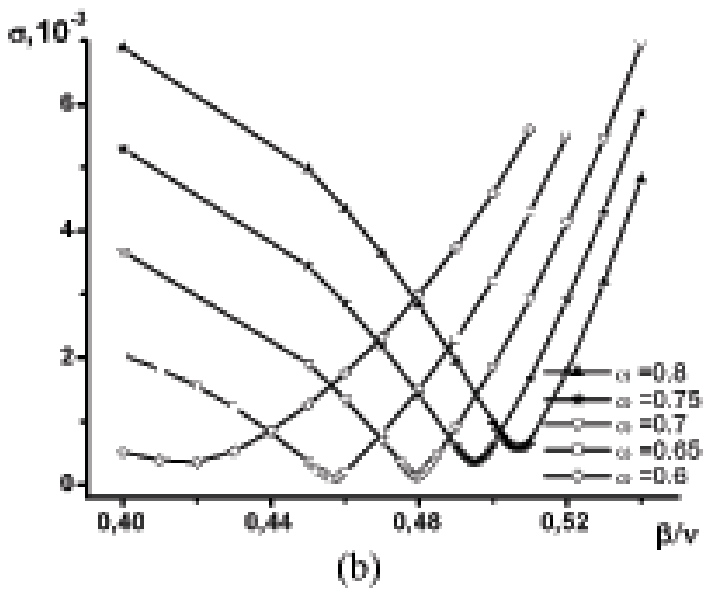}\\
\includegraphics[width=0.45\textwidth]{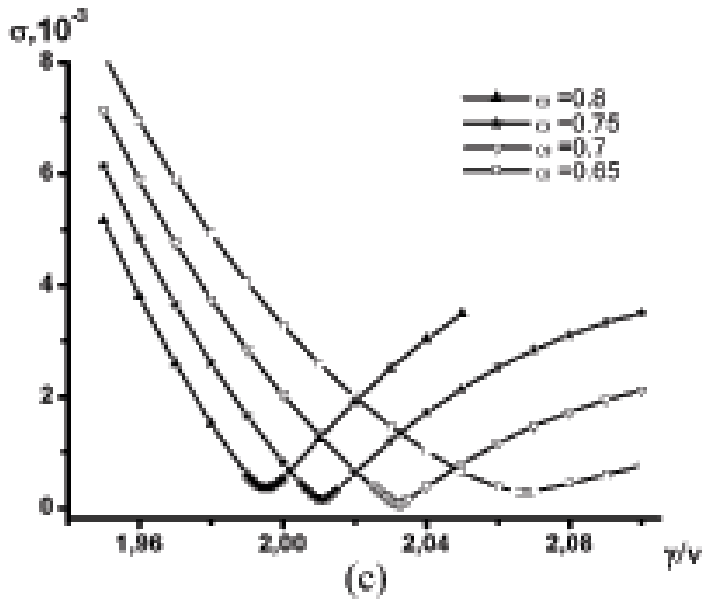}
\includegraphics[width=0.45\textwidth]{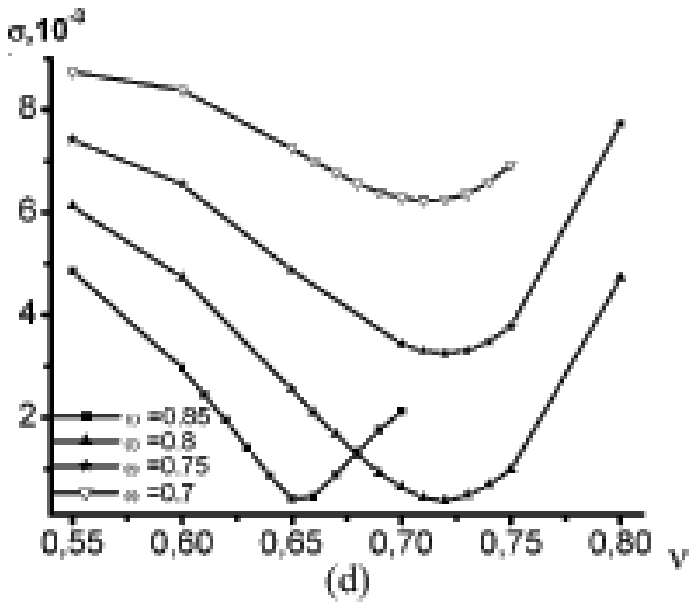}
\caption{ \label{fig:10} Dependence of the mean square errors $\sigma$ of the fits for
heat capacity (a), magnetization (b), susceptibility (c), and thermal
derivative of cumulant (d) as a function of the exponents $\alpha/\nu$,
$\beta/\nu$, $\gamma/\nu$, and $\nu$ for different values
of $\omega$. }
\end{center}
\end{figure}
\begin{eqnarray}
 C(L) &\sim& L^{\alpha/\nu} \left(1+a L^{-\omega}\right) \\
 M(L) &\sim& L^{-\beta/\nu} \left(1+b L^{-\omega}\right)   \\
 \chi(L) &\sim& L^{\gamma/\nu} \left(1+c L^{-\omega}\right)     \\
 \frac{dU}{dT}(L) &\sim& L^{1/\nu} \left(1+d L^{-\omega}\right)
\end{eqnarray}
\begin{table}[t]
\hspace*{-5mm}\parbox{0.47\textwidth}{
\caption{\label{tab:3} Values of the exponents $\alpha/\nu$, $\beta/\nu$, $\gamma/\nu$,
and $\nu$ with values of the exponent $\omega$, giving the best fit in
approximation procedure}
}
\hfill
\parbox{0.53\textwidth}{%
\caption{\label{tab:4} Values of the critical exponents obtained in present work for
average value of exponent $\omega =0.76$ and corresponding exponents from \cite{13}}} \\[2mm]
\hspace*{-3mm}\begin{tabular}{ccccc} \hline \hline
  {}          & {$\alpha/\nu$} & {$\beta/\nu$} & {$\gamma/\nu$} & {$\nu$}    \\ \hline
  {}          & $-0.096(3)$   & {$0.457(2)$}  & {$2.032(1)$}    & {$0.710(10)$} \\
  {$\omega$}  & $0.90$        & {$0.65$}      & {$0.70$}        & {$0.80$} \\ \hline
\end{tabular}
\hfill
\begin{tabular}{cllll} \hline \hline
  {}          & \multicolumn{1}{c}{$\alpha$}   & \multicolumn{1}{c}{$\beta$}
              & \multicolumn{1}{c}{$\gamma$}   & \multicolumn{1}{c}{$\nu$} \\ \hline
  {present}   & {$-0.078(30)$} & {$0.362(20)$} & {$1.441(15)$}  & {$0.710(10)$}    \\
  {\cite{13}} & {$-0.1048$}    & {$0.3504$}    & {$1.4453$}     & {$0.7155$} \\ \hline
\end{tabular}
\end{table}
makes it possible to determine the critical exponents
$\alpha$, $\nu$, $\beta$, $\gamma$, and $\omega$ by means of
statistical data processing of simulation results.
To analyze simulation data we have used the linear approximation
of the $(X L^{-\Delta})$ on $L^{-\omega}$ and then investigated the dependence
of the mean square errors $\sigma$ of this fitting procedure for the function
$X L^{-\Delta}(L^{-\omega})$ on the changing exponent $\Delta$ and $\omega$ values.
In Fig.~\ref{fig:10} we plot the $\sigma$ for heat capacity (Fig.~\ref{fig:10}a),
magnetization (Fig.~\ref{fig:10}b), susceptibility (Fig.~\ref{fig:10}c), and
temperature derivative of cumulant (Fig.~\ref{fig:10}d) as a function of the exponents
$\alpha/\nu$, $\beta/\nu$, $\gamma/\nu$, and $\nu$ for different values
of $\omega$. Minimum of $\sigma$ determines the values of exponents.
In Table~\ref{tab:3} we present the obtained values of the exponents $\alpha/\nu$,
$\beta/\nu$, $\gamma/\nu$, $\nu$, and $\omega$, which give minimal values
of $\sigma$ in these fits. Then we determine the average value of
$\omega=0.76(5)$ with the use of which there were computed the final values
of exponents. In Table~\ref{tab:4} there are presented the values of the
exponents obtained in this work by simulation methods and from \cite{13}
with the use of the field-theoretic approach and scaling relations for
critical exponents.

The comparison these values shows their good agreement within the
limits of statistical errors of simulation and numerical approximations
and good agreement with the values of the static critical exponents computed
by the short-time dynamics method.

\section{Measurements of the critical temperature and critical exponents
for 3D XY-model\label{sec:4}}

\begin{wrapfigure}[9]{r}{7cm}   
\vspace*{-3mm}\hspace*{-2mm}\includegraphics[width=0.45\textwidth]{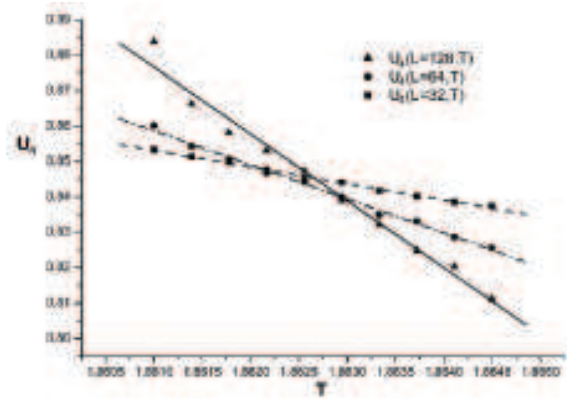}\vspace*{-12mm}\\
\caption{\label{fig:8} Binder cumulant $U_4(T,L)$ for 3D XY-model
as a function of T for lattices with different sizes $L$. }
\end{wrapfigure}
Also, we have carried out the Monte Carlo study of the effect of
LR-correlated quenched defects on the critical behavior of 3D XY-model
characterized by the two-component order parameter. As is well-known \cite{2,13},
renormalization group analysis predicts the possibility of a new type
of the critical behavior for this model different from the critical behavior
of pure XY-like systems or systems with point-like uncorrelated defects.
We considered the same site-diluted cubic lattices with linear
defects of random orientation in the samples with the spin concentration
$p=0.8$.
The critical temperature $T_c=1.8626(5)$ was determined by
the calculation of Binder cumulant $U_4(L,T)$ for lattices
with sizes $L$ from 32 to 128 (Fig.~\ref{fig:8}). For simulations we have used
the Wolff single-cluster algorithm.

\subsection{Evolution from an ordered state}\label{sec:4a}
\begin{wrapfigure}[24]{r}{7cm}   
\vspace*{-7mm}
\includegraphics[width=0.44\textwidth]{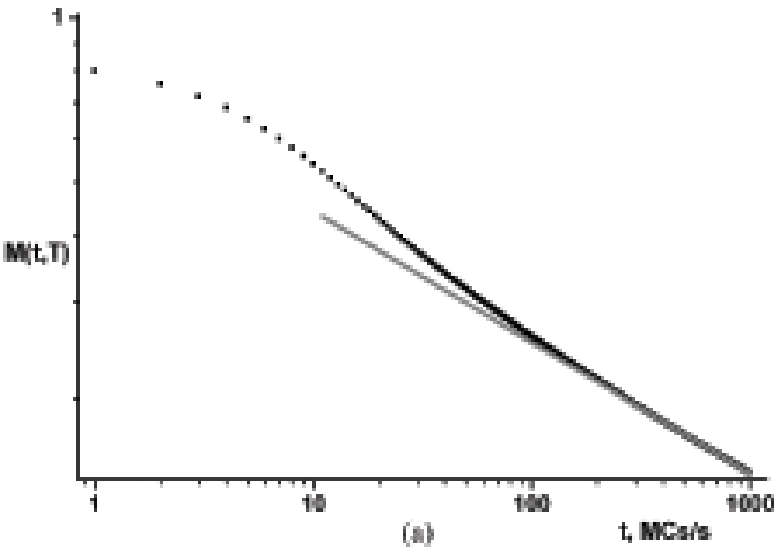}
\includegraphics[width=0.44\textwidth]{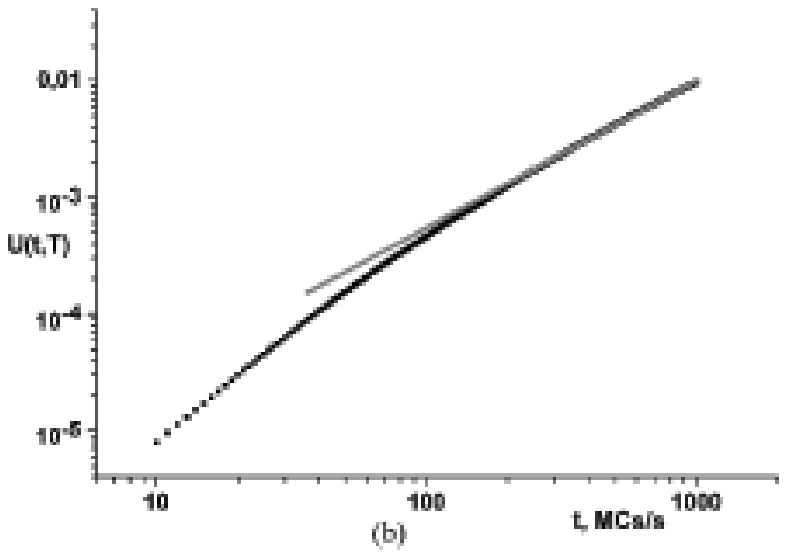}
\includegraphics[width=0.44\textwidth]{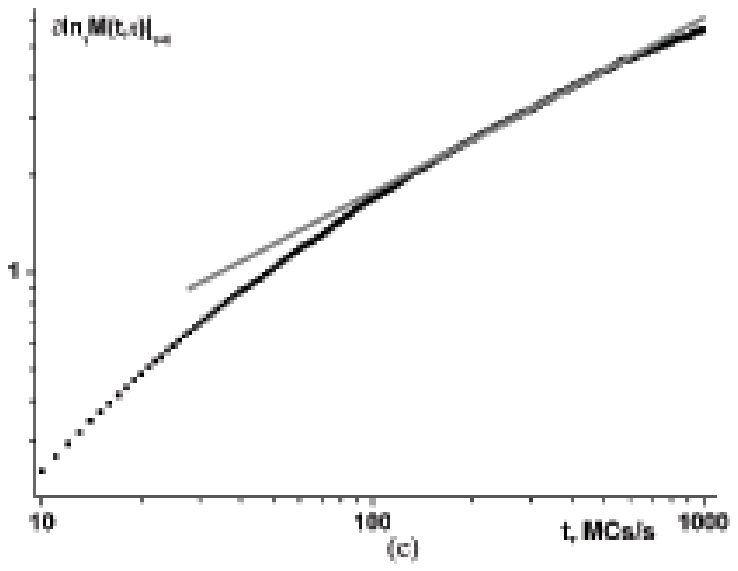}
\vspace*{-15mm}\\
\caption{ \label{fig:9} Time evolution of the magnetization $M(t)$ (a), Binder
cumulant $U_2(t)$ (b) and the logarithmic derivative of the magnetization
$\left.\partial_{\tau} ln  M(t,\tau) \right|_{\tau=0}$ (c)
for 3D XY-model with lattice size $L=128$ at the critical temperature $T_c=1.8626$.}
\end{wrapfigure}
We have performed simulations of the critical relaxation of the XY-model
with linear defects starting from an ordered initial state.
As example, in Fig.~\ref{fig:9} we show the obtained curves for the magnetization $M(t)$ (Fig.~\ref{fig:9}a),
Binder cumulant $U_2(t)$ (Fig.~\ref{fig:9}b) and the logarithmic derivative of the magnetization
$\left.\partial_{\tau} ln  M(t,\tau) \right|_{\tau=0}$ (Fig.~\ref{fig:9}c), which are plotted
in log-log scale up to $t=1000$ for lattices with $L=128$. These curves were resulted
by averaging over 3000 different samples.
On these figures we can observe the crossover from dynamics, which is similar to
that in the pure system on early times of the evolution up to $t=150$,
to dynamics of the disordered system with the influence of
linear defects in the time interval [350,800].
The slope of $M(t)$, $U_2(t)$ and $\left.\partial_{\tau} ln  M(t,\tau) \right|_{\tau=0}$
over the interval [350,800] provides the exponents $\beta/\nu z =0.239(1)$,
$d/z=1.221(21)$ and $1/\nu z =0.531(13)$, whereas our theoretical-field predictions \cite{13}
give the following values of exponents $\beta/\nu z =0.204$,
$d/z=1.268$ and $1/\nu z =0.556$.
As well as for Ising model, we have considered the corrections to the scaling in order
to obtain accurate values of the critical exponents in concordance with procedure,
which was discussed in Subsection~\ref{sec:3a}.

As a result of this analysis
we obtained the following values of critical exponents
\begin{equation}
\begin{array}{rl}
& z = 2.364 \pm 0.007,  \\
& \nu = 0.778 \pm 0.026,   \\
& \beta = 0.370 \pm 0.030,\\
& \omega = 1.05 \pm 0.04.
\end{array}
\end{equation}
The comparison of these values of exponents with those obtained in
\cite{13} with the use of the field-theoretic approach $z=2.365$,
$\nu=0.760$, $\beta=0.366$, and $\omega = 1.15$ in \cite{15} shows
their good agreement within the limits of statistical errors of
simulation and numerical approximations.

The obtained results confirm the strong effect of
LR-correlated quenched defects on both the critical behavior of 3D Ising
model and the systems characterized by the many-component order parameter.

\subsection{Evolution from a disordered state}\label{sec:4b}
\begin{figure}
\includegraphics[width=0.45\textwidth]{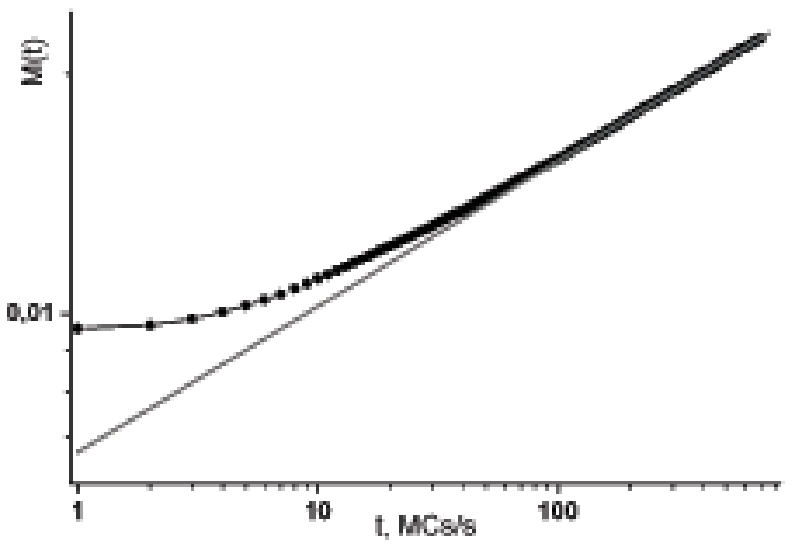}
\includegraphics[width=0.45\textwidth]{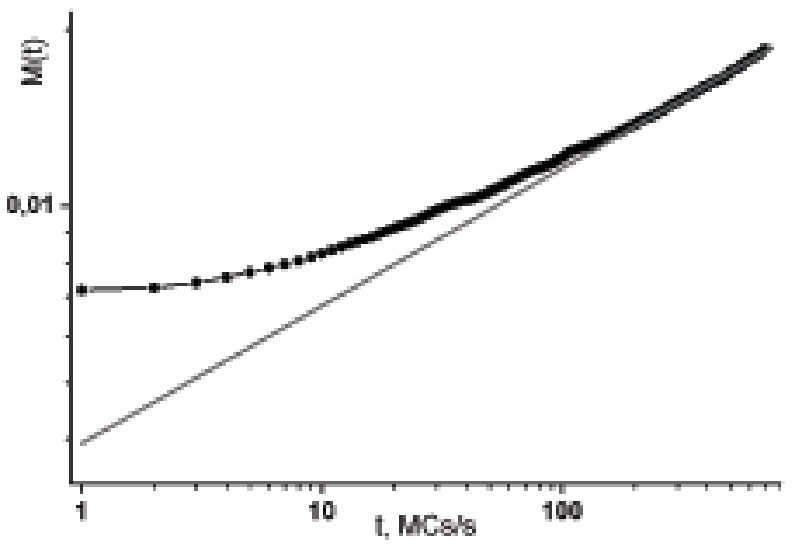}\\
\includegraphics[width=0.45\textwidth]{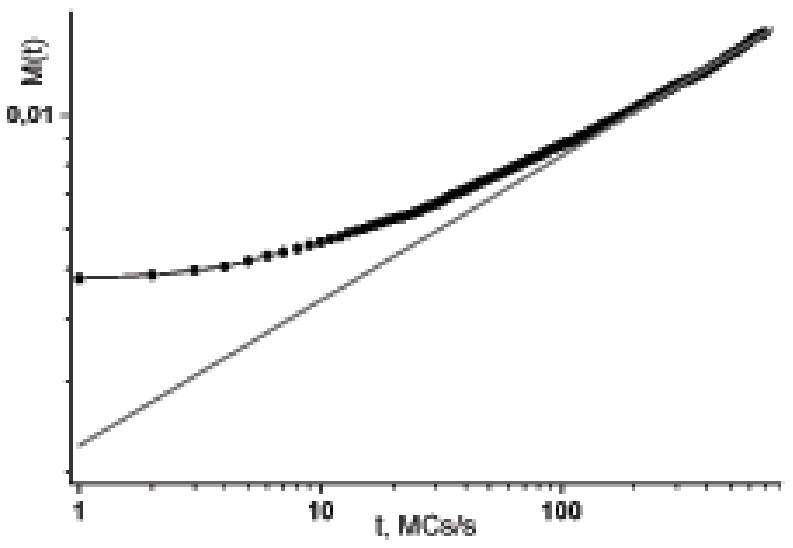}
\includegraphics[width=0.45\textwidth]{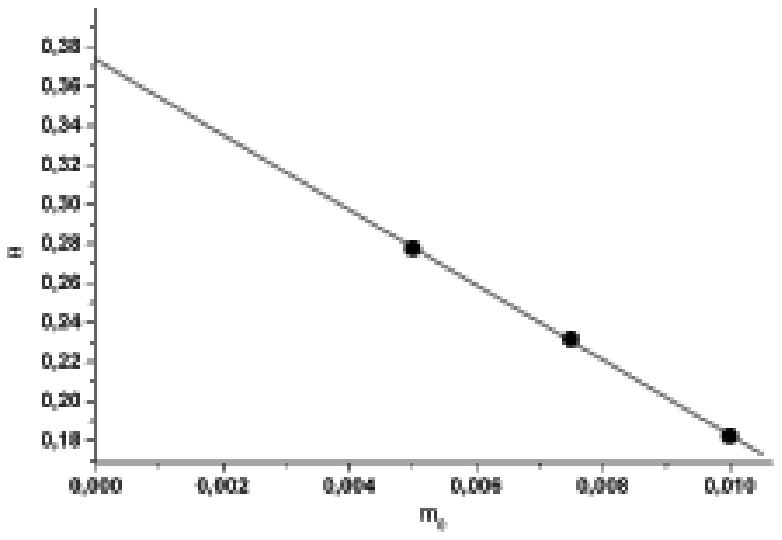}
\caption{ \label{fig:11} Time evolution of the magnetization $M(t)$
for 3D XY-model with the initial magnetization $m_0=0.01$ (a), $m_0=0.0075$ (b) and
$m_0=0.005$ (c) at the critical temperature $T_c=1.8626$; determination of
asymptotic value of the dynamic critical exponent $\theta$ in the limit $m_0\to 0$ (d).}
\end{figure}

We have also performed simulations of evolution of the system with linear defects
on the largest lattice with $L=128$, starting from a disordered state with
small initial magnetizations $m_0=0.01$, $m_0=0.0075$ and $m_0=0.005$ at the critical point.
The initial magnetization has been prepared by flipping in an ordered state
a definite number of spins at randomly chosen sites  in order to get the desired
small value of $m_0$. In accordance with Section~\ref{sec:2}, a generalized dynamic
scaling predicts in this case a power law evolution for the magnetization $M(t)$,
the second moment $M^{(2)}(t)$ and the auto-correlation $A(t)$ in the short-dynamic region.

\begin{figure}
\centerline{
\includegraphics[width=0.5\textwidth]{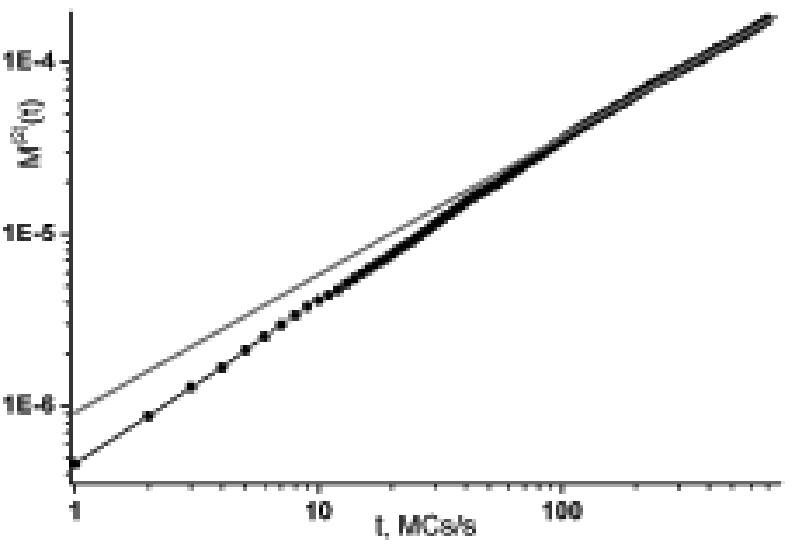}
\includegraphics[width=0.5\textwidth]{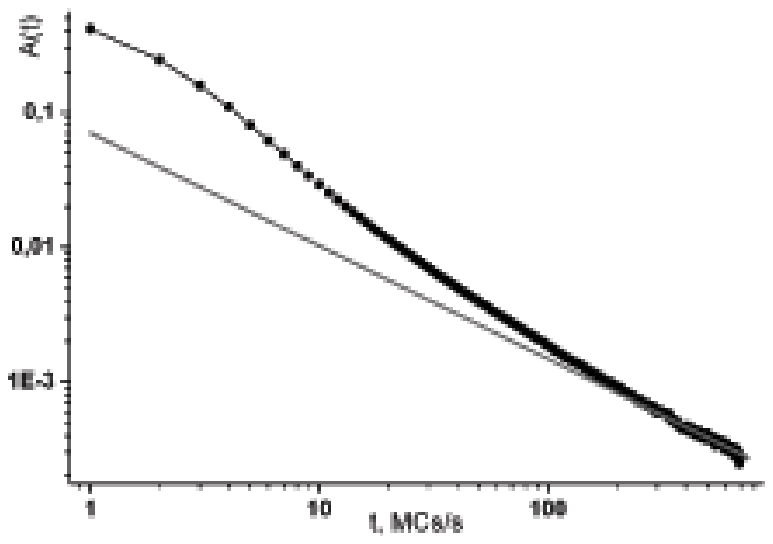}}
\caption{ \label{fig:12} Time evolution of the second moment $M^{(2)}$ (a)
and the auto-correlation $A(t)$ (b) for 3D XY-model with the initial magnetization $m_0=0$
at the critical temperature $T_c=1.8626$. }
\end{figure}

In Fig.~\ref{fig:11} and \ref{fig:12} we show the obtained curves for $M(t)$ (Fig.~\ref{fig:11}),
$M^{(2)}(t)$ and $A(t)$ (Fig.~\ref{fig:12}), which are plotted
in log-log scale up to $t=700$. These curves were resulted by averaging
over 3000 different samples with 25 runs for each sample.
From Fig.~\ref{fig:11} we can see also an initial increase of the magnetization,
which is a very prominent phenomenon in the short-time critical dynamics.
But in contrast to dynamics of the pure systems, we can observe,
as previously for Ising model, the crossover from dynamics of the pure system
on early times of the magnetization evolution up to $t=100$
to dynamics of the disordered system with the influence of
linear defects in the time interval [200,650]. In this time interval
we determined the values of the dynamic critical exponent $\theta$,
which are equal $\theta=0.182(3)$ for case with the initial magnetization $m_0=0.01$,
$\theta=0.232(7)$ for $m_0=0.0075$ and $\theta=0.278(11)$ for $m_0=0.005$.
Then, in accordance with (\ref{eq:theta}) the asymptotic value of the exponent
$\theta=0.374(14)$ was determined in the limit $m_0\to 0$ (Fig.~\ref{fig:11}d)
with the use of linear extrapolation.

The analysis of results for evolution of the second moment $M^{(2)}(t)$
(Fig.~\ref{fig:12}a) and the autocorrelation $A(t)$ (Fig.~\ref{fig:12}b),
obtained for simulation of systems with the initial magnetization $m_0=0$
(in reality for $m_0=10^{-6}$ such as for XY-model the spin
configuration with $m_0=0$ is impossible to prepare), gives directly the
values of exponents $c_2=0.825(23)$ and $c_a=0.907(30)$ for the time
interval [300,650]. The same crossover phenomena is observed in evolution
of $M^{(2)}(t)$ and $A(t)$ from dynamics of the pure system
on early times to dynamics of the disordered system with the influence of
linear defects.

On basis of these values of exponents $\theta$, $c_2$ and $c_a$
we obtained the exponents $z=2.342(57)$ and $\beta/\nu=0.534(35)$, which
quite well agree with results of simulation from an ordered
initial state with $m_0=1$ and with results of theoretical field description
for XY-like systems with linear defects \cite{13} within the
limits of statistical errors of simulation and numerical approximations.

\section{Conclusion remarks\label{sec:5}}

The present results of Monte Carlo investigations allow us to recognize that
the short-time dynamics method is reliable for the study of the critical
behavior of the systems with quenched disorder and is the alternative to
traditional Monte Carlo methods. But in contrast to studies of the critical
behavior of the pure systems by the short-time dynamics method, in case of the
systems with quenched disorder corresponding to randomly distributed linear defects
after the microscopic time $t_{mic}\simeq 10$ there exist three stages of dynamic
evolution. For systems starting from the ordered initial states ($m_0=1$)
in the time interval of 50-200 MCS, the power-law dependences are observed
in the critical point for the magnetization $M(t)$, the logarithmic derivative
of the magnetization $\left.\partial_{\tau} ln  M(t,\tau) \right|_{\tau=0}$ and
Binder cumulant $U_2(t)$, which are similar to that in the pure system.
In the time interval [450,900], the power-law dependences are observed in the
critical point which are determined by the influence of disorder.
However, careful analysis of the slopes for $M(t)$,
$\left.\partial_{\tau} ln  M(t,\tau) \right|_{\tau=0}$ and $U_2(t)$ reveals that
a correction to scaling should be considered in order to obtain accurate
results. The dynamic and static critical exponents were computed with the
use of the corrections to scaling for the Ising and XY models with linear
defects, which demonstrate their good agreement with results of the
field-theoretic description of the critical behavior of these models
with long-range correlated disorder. In intermediate time interval of
200-400 MCS the dynamic crossover behavior is observed from the critical
behavior typical for the pure systems to behavior determined by the
influence of disorder.

The investigation of the critical behavior of the Ising model with extended defects
starting from the disordered initial states with $m_0 \simeq 0$ also have
revealed three stages of the dynamic evolution. It was shown that
the power-law dependences for the magnetization $M(t)$,
the second moment $M^{(2)}(t)$ and the autocorrelation $A(t)$
are observed in the critical point, which are typical for the pure system
in the interval [10,70] and for the disordered system in the interval [100,650].
In intermediate time interval the crossover behavior is observed in the dynamic
evolution of the system.
The obtained values of exponents demonstrate a good agreement within the
limits of statistical errors of simulation and numerical approximations with
results of simulation of the pure Ising model by the short-time dynamics method \cite{13}
for the first time interval and with our results of simulation of the critical
relaxation of this model from the ordered initial states.

Also, we would like to note that over complicated critical dynamics of the systems
with quenched disorder the accurate determination of the critical temperature
is better to carry out in equilibrium state from the coordinate of the points
of intersections of the curves specifying the temperature dependence of Binder
cumulants $U_4 (L,T)$ or ratio $\xi/L$ for different linear sizes $L$ of lattices.

The obtained results for 3D XY-model confirm the strong
influence of LR-correlated quenched defects on the critical behavior of
the systems described by the many-component order parameter.
As a result, a wider class of disordered systems,
not only the three-dimensional Ising model, can be characterized by a new type
of critical behavior induced by quenched disorder.

We are planning to continue the Monte Carlo study of critical behavior
of the model with LR-disorder for different spin concentrations $p$
and investigate the universality of critical behavior of diluted
systems with LR-disorder focusing on the problem of disorder independence
of asymptotic characteristics.

\section*{Acknowledgements}
The authors would like to thank Prof.~N.Ito, Prof.~J.Machta and Prof.~W.Janke
for useful discussion of results of this work during The 3-rd International Workshop
on Simulational Physics in Hangzhou (November 2006).
This work was supported in part by the Russian Foundation for Basic Research
through Grants No. 04-02-17524 and No. 04-02-39000,
by Grant No. MK-8738.2006.2 of Russian Federation President
and NNSF of China through Grant No.~10325520.

%

\end{document}